\documentclass[conference]{IEEEtran}
\IEEEoverridecommandlockouts
\usepackage{amsmath,amssymb,amsfonts}
\usepackage{algorithmic}
\usepackage{graphicx}
\usepackage{todonotes}
\usepackage{multirow}
\usepackage{textcomp}
\usepackage{biblatex}
\usepackage{footmisc}

\usepackage{booktabs}
\usepackage{siunitx}
\usepackage[table]{xcolor}
\usepackage{xparse}
\usepackage{booktabs,adjustbox,caption,array}

\usepackage{xcolor}
\def\BibTeX{{\rm B\kern-.05em{\sc i\kern-.025em b}\kern-.08em
    T\kern-.1667em\lower.7ex\hbox{E}\kern-.125emX}}

\usepackage[most]{tcolorbox}

\definecolor{beige}{HTML}{F5F5DC}

\begin{document}

\title{From LLMs to Agents in Programming: The Impact of Providing an LLM with a Compiler}



\author{%
\IEEEauthorblockN{Viktor Kjellberg, Miroslaw Staron, Farnaz Fotrousi}
\IEEEauthorblockA{%
\textit{Computer Science and Engineering}\\
\textit{Chalmers University of Technology}\\
\textit{and University of Gothenburg}\\
viktor.kjellberg@gu.se, miroslaw.staron@gu.se, farnaz.fotrousi@gu.se
}%
}

\maketitle

\begin{abstract}

Large Language Models have demonstrated a remarkable capability in natural language and program generation and software development. However, the source code generated by the LLMs does not always meet quality requirements and may fail to compile. Therefore, many studies evolve into agents that can reason about the problem before generating the source code for the solution. The goal of this paper is to study the degree to which such agents benefit from access to software development tools, in our case, a \texttt{gcc} compiler. We conduct a computational experiment on the RosettaCode dataset, on 699 programming tasks in C. We evaluate how the integration with a compiler shifts the role of the language model from a passive generator to an active agent capable of iteratively developing runnable programs based on feedback from the compiler. We evaluated 16 language models with sizes ranging from small (135 million) to medium (3 billion) and large (70 billion). Our results show that access to a compiler improved the compilation success by 5.3 to 79.4 percentage units in compilation without affecting the semantics of the generated program. Syntax errors dropped by 75\%, and errors related to undefined references dropped by 87\% for the tasks where the agents outperformed the baselines.    
We also observed that in some cases, smaller models with a compiler outperform larger models with a compiler. We conclude that it is essential for LLMs to have access to software engineering tools to enhance their performance and reduce the need for large models in software engineering, such as reducing our energy footprint. 


\end{abstract}


\begin{IEEEkeywords}
AI Agents, LLM, Compiler, Program Synthesis
\end{IEEEkeywords}

\section{Introduction}
\label{sec:introduction}
\noindent

Software development has undergone rapid changes in the last few years, driven by the development of Large Language Models (LLMs) for program synthesis \cite{Xinyi_Hou_et_al, peng2023the}. LLMs have demonstrated an impressive ability to generate code from natural language instructions by prompting, which has led to their wide implementation as tools for automated test improvements \cite{automatedunittestimprovement} and for code generation \cite{Isyourcodecorrect, chen2021evaluatinglargelanguagemodels}. However, using LLMs for code synthesis has some limitations, the most significant of which is the unidirectional flow from prompt to model to generated output. 
This means that the generation is based on neural networks and does not guarantee the correctness of the generated code \cite{jesse2023largelanguagemodelssimple}, nor does it ensure structural and functional quality. The primary approach to addressing this issue is to fine-tune LLMs on code, increase the model size, and provide extra context from relevant sources, thereby enhancing their capabilities. However, increasing the size and training data of large models requires a significant amount of resources and effort.  
Fine-tuning models on code can enhance a model's capability for program synthesis \cite{rozière2024codellamaopenfoundation, CYCLE}. However, this process still requires a significant amount of computational resources \cite{strubell2020energy}. Using Retrieval-Augmented Generation (RAG) to retrieve relevant context for the LLM has also been shown to significantly increase the model's performance in solving programming-related tasks \cite{wang2025coderagbenchretrievalaugmentcode}. Even though increasing model and data size, fine-tuning models or providing extra context can lead to improved performance, it does not solve the underlying problem of the lack of quality assurance, without which we cannot guarantee that the code is even executable.

The one-way process from input to the generated program gives the LLM only one chance to solve the tasks at hand, with the initial context and the task as the sole references. Even with approaches like RAGs or Chain of Thought (CoT), the generation is based on creating new tokens without an oracle to determine whether the system actually works. If the problem description is clear and the task is simple, with similar tasks available in the training data for the LLM, this might be sufficient. However, we do not need LLMs that simply replicate existing solutions; we need ones that solve new problems. Even reasoning models (like GPT-4o, GPT-5, DeepSeek-R1) utilize advanced strategies like Mixture-of-Experts to improve their problem-solving ability. Therefore, we need to design program generation systems (Agents) that combine generation with the ability to validate the solutions. 

LLMs-based agents can combine multiple agents with different roles \cite{lin2024soen101codegenerationemulating}, such as developer, tester, and debugger, or integrate external tools \cite{9216382, bi-etal-2024-iterative, zhang-etal-2024-codeagent, grubisic2024compilergeneratedfeedbacklarge}.
Integrating external tools, such as compilers, with the agents can enhance the quality of output generated by a single LLM through collaboration and feedback \cite{bi-etal-2024-iterative}. 
However, in these cases, the agent also has access to other tools, such as databases \cite{bi-etal-2024-iterative} or web searches \cite{hu-etal-2025-compileagent}.

Therefore, in our first research question, we analyzed the significance of agent systems with Language Models (LMs) that interact with a compiler in relation to the success rate of compiling the generated programs:

\textit{RQ1: To what extent does granting an LM (of size 135 million to 70 billion parameters) access to a compiler increase the rate of generating executable programs compared to a model-only baseline?}

Our goal in this study was not to introduce a new framework, but rather to examine the interaction between a single-model agent and a compiler. We evaluated LMs of varying sizes (135 million to 70 billion parameters) on a broad range of programming tasks. However, the size of the model was not the only parameter that may have affected the success rate of compiling the code; the type of problem may also have had an impact. Programming tasks can vary in complexity, and the feedback from the compiler may have had a different effect on the generated solutions depending on the task. So, we analyzed even that:

\textit{RQ2: Which patterns can be identified in the tasks that benefit the most when an LM has access to a compiler?}


We observed in previous work \cite{bi-etal-2024-iterative} that the overall rate of errors should be reduced by providing the LLM with feedback from the compiler. However, compiler errors can, in some cases, be misleading about the actual cause of the error. If we compile a file containing only text without any code, we can assume that we will get a syntax error from the compiler. But categorizing this as simply a syntax error would be misleading because the actual error is not the wrongly formatted code, but rather the lack of code. When the model receives feedback from the compiler, it may therefore misinterpret it. It was thus essential to know what type of error the integration of a compiler might have helped the LM to solve and which errors it did not help to solve:

\textit{RQ3: Which patterns can be identified in the errors that benefit the most when an LM has access to a compiler?}


To address these questions, we conducted a computational experiment with 16 different LMs. We studied whether the generated programs compile, and how similar they were to the original RosettaCode implementation in terms of textual similarity (BLEU, ROUGE), and semantics (CodeBERTScore). We also analyzed the patterns in the results to address the last two research questions. In our study, we evaluated both the models and the programs to understand the patterns in both. 

Our results show that providing a LM access to a compiler increased the rate of generating executable programs, and the effect varied between the models. However, the size of the model did not correlate with the increased rate of executable programs generated by the models, and both small and large models saw an increase. The semantic and textual similarity between the generated programs from agents and baselines to the original RosettaCode was similar. Therefore, the feedback from the compiler enhanced the model's ability to generate executable programs by resolving errors without altering the program's semantics. 

The compiler helps the agents solve syntax and undefined references, reducing these by 75\% and 87\%. What type of task was, however, not of any significance for the agent's ability to solve the tasks.

Our paper is structured as follows. Section \ref{sec:related_work} reviews the most significant related work in this area. In Section \ref{sec:methodology} the research method and evaluation metrics used are presented. The findings of the experiment are then presented in Section \ref{sec:result}. Lastly, the threats to validation are presented in Section \ref{sec:threats} followed by a summarizing conclusion in \ref{sec:conclusion}.

\section{Related Work}
\label{sec:related_work}

There exist many frameworks for agents to solve programming-specific tasks where the agent iteratively works their way toward a solution with the help of external tools \cite{li2024codetreeagentguidedtreesearch, bi-etal-2024-iterative, ishibashi2024selforganizedagentsllmmultiagent,hu-etal-2025-compileagent,9216382, zhang-etal-2024-codeagent, grubisic2024compilergeneratedfeedbacklarge, CYCLE}. 

Many frameworks use discussion among agents, where the agents are assigned different roles in the process. It can take the form of a discussion between high-level planners and developers \cite{li2024codetreeagentguidedtreesearch, ishibashi2024selforganizedagentsllmmultiagent, 9216382} where one agent instructs the other by transforming the task into smaller actions.

A common feature of these frameworks is the inclusion of an agent responsible for the quality of the generated output. In \cite{ishibashi2024selforganizedagentsllmmultiagent, li2024codetreeagentguidedtreesearch, 9216382} they include a agent that can execute generated or predefined unit tests. However, generated unit tests need to be evaluated independently, and using predefined tests limits the scope of tasks to which the agents can be applied. 

Instead of solely relying on unit tests, some frameworks include a compiler with the agents, which allows the agents to make decisions based on the feedback from the compiler. Jierui Li et al. \cite{li2024codetreeagentguidedtreesearch} propose a framework consisting of a Thinker agent that develops a high-level strategy, which is then given to the Solver agent that generates a set of initial solutions. The solution is iteratively passed to the Thinker agent to improve the generated solution together with a Debugger agent. Where the Debugger agent has access, among other things, to a compiler. They evaluated their models with pass@1 on a wide range of benchmarks, such as HumanEval and MBPP, but they did not, however, evaluate the effect of adding the compiler specifically, but rather the whole framework.

Another framework with a compiler as an essential part is CoCoGen presented in \cite{bi-etal-2024-iterative}. Here, they evaluate their framework for solving tasks in a project repository, where the agent can retrieve information about functions and extract code to address the current problem. The LLM then generates code based on the retrieved context and tests it through a compiler. If the compiler fails, the error message is used to retrieve new information that might help the LLM to resolve the error. In this paper, Zhangqian Bi et al. show that the compiler alone did improve the result and that the improvement peaked at the third iteration. However, they only evaluate their framework on GPT-3.5-Turbo and Code Llama on programs written in Python at the project level.

Grubisic, Dejan, et al. \cite{grubisic2024compilergeneratedfeedbacklarge} evaluate the use of a fine-tuned Llama 2 model for optimizing IR for LLVM assembly. Their framework follows a similar process to the one presented in this paper. However, the agent is asked to optimize a given code and then generate the appropriate passes for compilation. They show that the results improve by allowing the model to iterate, but that the improvement plateaus after a few iterations. They also show that the performance of the agent improves after feedback from the compiler.

\begin{figure*}[ht]
  \includegraphics[width=17cm]{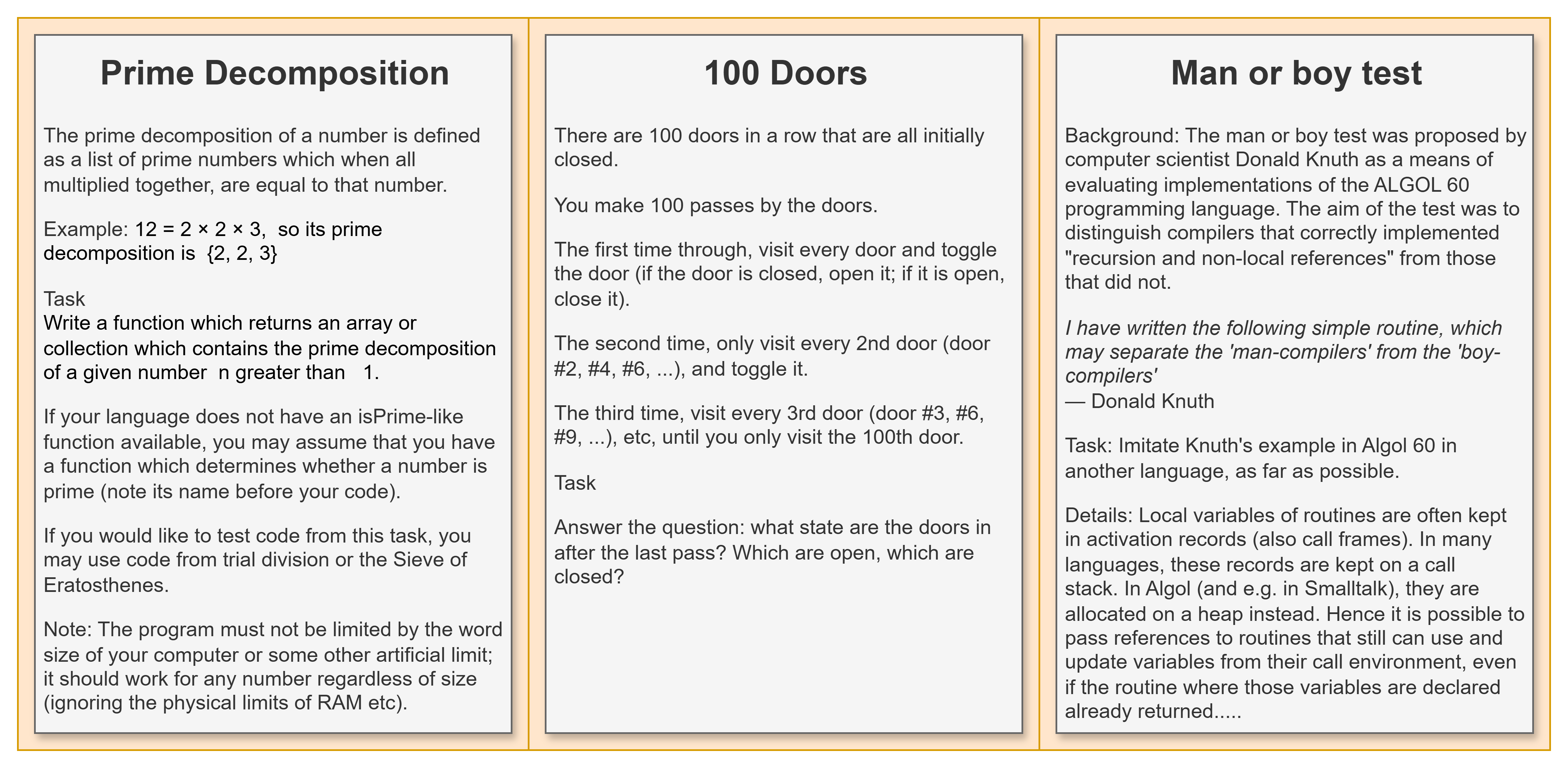}
  \caption{Three examples of tasks found in RosettaCode}
  \label{fig:RosettaCodeEx}
\end{figure*}

The LMs used in these studies have been models with 7 billion parameters or larger. However, the effect of the model's size on its performance remains an unexplored area. If we could use smaller models and still achieve the same rate of successful compiled code on the generated programs, we could run agents on less resource-heavy hardware.

\section{Methodology}
\label{sec:methodology}
\noindent

We conducted an exploratory study of 16 models in a simplified agent framework where the LM interacts with a compiler to generate code in C. We analyze whether access to a compiler can improve the capability of smaller models to a greater degree than that of larger models. Therefore, the chosen models were selected to cover a range of different sizes, spanning from 135 million to 70 billion parameters.  
We used C code in our experiments for two reasons: it utilizes a compiler and is a common language in embedded systems, which is essential for our industrial partners.

We used a dataset consisting of a wide range of tasks to evaluate the agent's capability to generate executable code and assessed the code based on textual and semantic similarity to the ground-truth solutions. We also analyzed the type of errors and tasks that the agents were able to solve to a higher degree than the baselines.

\subsection{RosettaCode} 

\begin{figure*}[t]
  \includegraphics[scale=0.5]{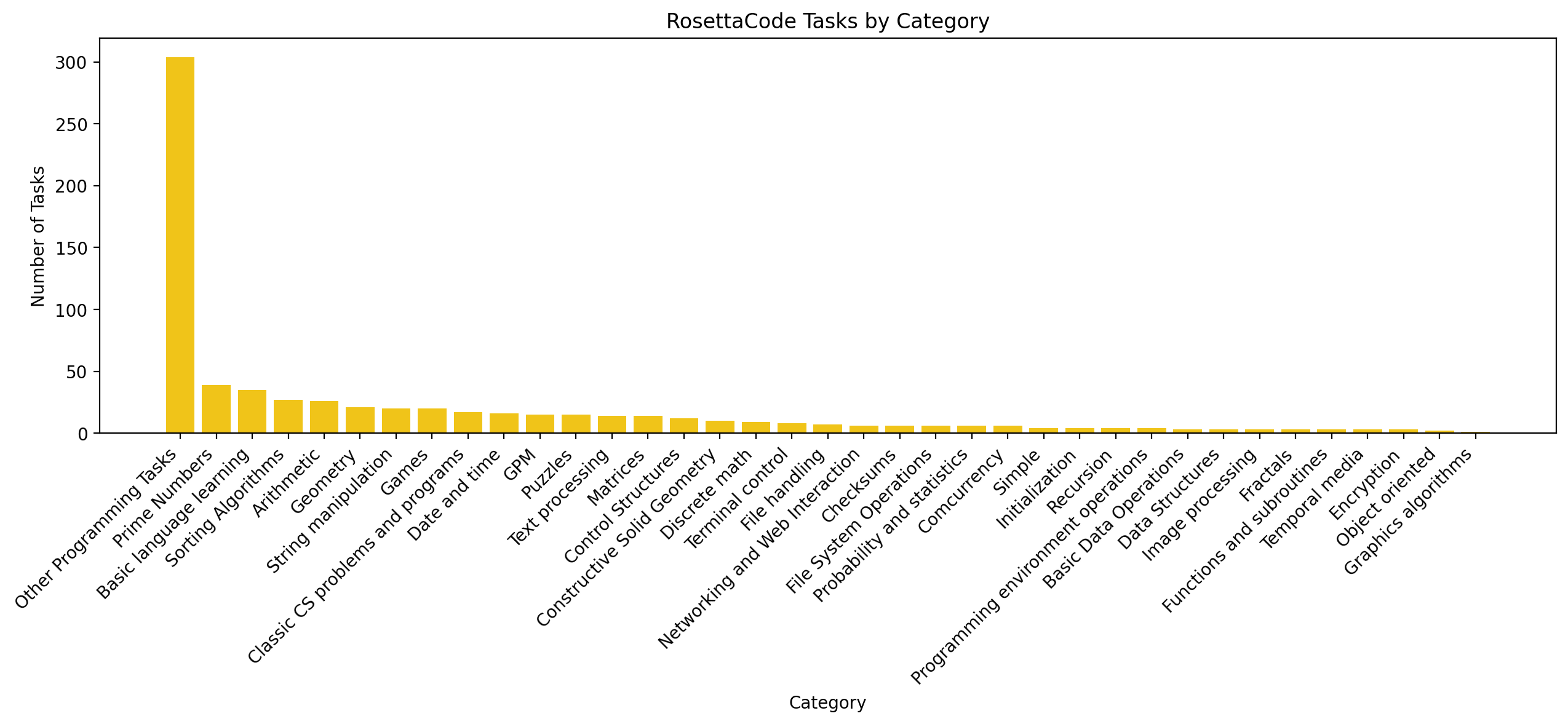}
  \caption{The number of tasks in each category in our subset of RosettaCode}
  \label{fig:RosettaCodeCat}
\end{figure*}

RosettaCode\footnote{https://github.com/acmeism/RosettaCodeData/tree/cb74b7914} is a programming chrestomathy consisting of 1,333 tasks with solutions in up to 976 different programming languages. The dataset contains a range of tasks of varied complexity, from simpler programming tasks that can be completed with a single function, e.g., a Hello World program, to more complex problems, such as 3D animations, games, and complex sorting algorithms. See Figure \ref{fig:RosettaCodeEx} for three examples of tasks that can be found in RosettaCode.
Each task includes the task name, a detailed description of its objective, and an executable solution in one or more of the 976 supported languages. The wide range of supported languages, which includes C, makes this dataset suitable for evaluating languages not found in other standard benchmarks, such as HumanEval \cite{chen2021evaluatinglargelanguagemodels} or MBPP \cite{austin2021programsynthesislargelanguage}. The large number of tasks and the variation make it suitable as a dataset for evaluating real-world programming tasks.  

However, not all tasks in the RosettaCode dataset have solutions written in C. The tasks used in this paper consisted of a subset of the tasks found in RosettaCode and were chosen based on two criteria. Each task must have a solution written in C and be compiled as is. Each solution written in C was therefore compiled, and if successful, added to the subset. Based on these criteria, a subset of 699 tasks with descriptions and solutions was chosen. Every task in RosettaCode is included in at least one of the 63 task categories found on the website for RosettaCode \cite{rosettacodeCategorySolutionsProgramming}.
Some categories have multiple sub-categories, and some tasks belong to multiple categories. But for our subset of tasks, each task was linked to a single category based on its highest relevance. The categories for our 699 tasks can be seen in Figure \ref{fig:RosettaCodeCat}. 

Our subset of RosettaCode, together with the following evaluation, was made public for review\footnote{https://doi.org/10.5281/zenodo.17361190}.

\subsection{Baselines}

We utilized 16 LMs from various families and sizes to establish a baseline for the subsequent experiment. Each LM was given the following role description as part of the prompt: \textit{You are a software that writes C programs based on prompts. Provides only the code, with no description}. This instructed the LM to act as desired, generating the answer in code without any extra descriptions that could disrupt the process of extracting the code from the output. It also instructed the model to only give answers in the desired programming language, C. The resulting output was then evaluated using the semantic and textual similarity to the ground-truth solution. The generated code was also assessed by compiling the extracted markdown code from the generated output. Each task was given once to each model, and they had one attempt to generate the code. We made the repository\footnote{https://anonymous.4open.science/r/ChatAI-845D} with the code for both running the baselines and the agent and evaluation available for review.

\subsection{Agent framework}

\begin{figure*}[t]
\centering
  \includegraphics[width=18cm]{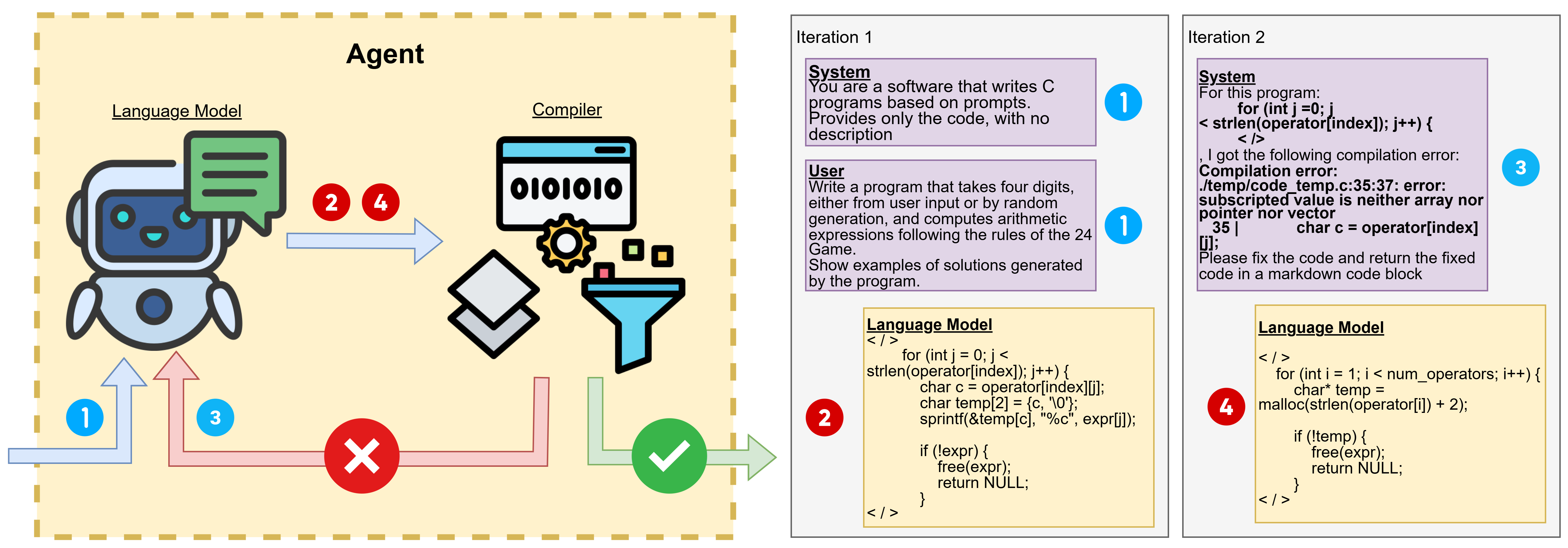} 
  \caption{The flow of content through the Agent system from the task description to the LM to the compiler (left). An example conversation with two iterations with the agent (right). Where the first iteration starts with the role description given by the system, followed by the task description and the generated program from the LM. The second iteration starts with the description of the errors given by the compiler. Each interaction and related step in the agent flow is mapped with a number. (1) The initial role and task description, (2, 4) the generated program from the LM, (3) feedback from the compiler with an error message.}
  \label{fig:agentFramework}
\end{figure*}

The agent followed our own architecture, which allowed us complete control over the model's configurations. The architecture can be seen in Figure \ref{fig:agentFramework} and consists of a generative LM serving as the generative engine, generating code suggestions based on a given task description. The code suggestion was then compiled with a \texttt{gcc} compiler, and if successful, the process ended. However, if the compiler returned an error, it was aggregated with the original task description and the faulty code and fed back to the LM. The LM was then given another attempt to generate code, using the error message from the previous compiler as context. This process continued until either the generated code successfully compiled or the maximum number of five iterations was met. To ensure the generated output from the LM was in the correct format, it was instructed with the same prompt as the baselines for the first iteration, assuming the role of software that produces code in the programming language C. In all following iterations, the following prompt was used: \textit{For this program \{CODE\}, I got the following compilation error: \{ERROR\}. Please fix the code and return the fixed code in a markdown code block.}, where \{CODE\} was the generated program from the previous iteration, and \{ERROR\} was the compiler error message for that program. In each iteration, the LM-generated answer and the compiler's error log were aggregated into a single log that served as the agent's internal memory. For each task, the log was saved and wiped, and a new instance of the agent was created.

\subsection{Dependent variables and analysis methods} 
\label{sec:Metrics}

To answer RQ1, we evaluated the generated programs from the agents based on five metrics and compared the results with those of the baselines.  

\textbf{Successful Compile Result} refers to the percentage of generated code that was compiled without any errors. 

\textbf{BLEU} and \textbf{ROUGE} measure the textual similarity between a generated program and the ground-truth solution, and were used to evaluate the difference in similarity between the generated code from both the agents and the baselines, and the ground-truth solutions. BLEU is a metric that rewards safe implementations without additional code that could be wrong. However, the generated code can differ from the ground-truth solution, but still be semantically correct. Therefore, ROUGE 1-gram, which instead rewards if the generated code includes all the relevant parts from the ground-truth solution, can be found in the generated solution.

In addition to measuring textual similarity, we also evaluate semantic similarity using \textbf{CodeBERTScore} \cite{zhou2023codebertscoreevaluatingcodegeneration} precision and recall. Where the embedded layers are compared in a BERT model, fine-tuned on programs written in C, for the generated program and the ground-truth solution.

To answer the second and third research questions, we shifted our focus from the performance of the models to their combined performance on each task.

For RQ2, we analyzed whether the task category, the amount of context provided in the task description, and the expected size of the generated solutions affected the agents' success rate. The context and expected size of the generated solution were defined by the number of tokens found in the task descriptions and the ground-truth solutions. The Pearson correlation was calculated between both the token length of the task descriptions and the ground-truth solutions, and the difference in the number of agents and baselines that solved the tasks.

To evaluate the result for RQ3, the generated output that failed to compile was annotated based on the following criteria: firstly, whether the generated output included any code, secondly, which programming language the code was written in, and thirdly, if the generated output included both code and text, whether the code was correctly indicated. The algorithm used for extracting code searched for code found between two sets of backticks (``` code ```). Only the first instance was used if multiple sets of backticks were found. If the algorithm did not find any match, the whole output was assumed to consist solely of code. 

After the generated output was labeled, the compiler's error messages were labeled based on the most common error.

\section{Results}
\label{sec:result}


\subsection{To what extent does granting a LM (of
size 135 million to 70 billion parameters) access to a compiler
increase the rate of generating executable programs compared
to a model-only baseline?}

Table \ref{tab:allresults} shows the increased rate of successfully compiled generated programs by each agent compared to the baseline. The performance of the baselines varied between types and sizes, and the percentage of successful compilations of generated code ranged from 2.9\% to 92.0\%. Here, the smallest model in our study, SmolLM 2, with 135 million parameters, achieved the lowest number of successful compilations (2.9\%), and Llama 3.3, with 70 billion parameters, achieved the highest (92.0\%). The size of the model had a significant Pearson correlation coefficient of 0.66 with the success rate of compiled programs.

The rate of successfully compiled programs increases when examining the agents performance. Access to the compiler had a positive effect, with an increase of 5.3 to 79.4 percent units, depending on the model. Llama 3.3 was able to increase the success rate of compiled programs to 99.9\% with only one failed task. The largest effect could be found for both of the Qwen 3 models. Qwen 3, with 4 billion parameters, increased its success rate by 79.4 percentage points in a successfully compiled code, ultimately achieving a success rate of 97.4\%. Qwen 3, with 30 billion parameters, increased its success rate by 33 percentage points, making both models achieve a success rate above 90\%. This is a notable difference from their respective baseline, where they performed worse than other similarly sized models.  
The baseline model for Qwen 3 was only able to outperform SmolLM 2 with 135 and 350 million parameters, while as an agent, it was among the top five best-performing ones.

Although the percentage of successfully compiled code measures how well the models can generate executable code, i.e., code without syntax errors, undeclared variables, linker errors, or type errors, we needed to analyze the scores measured with ROUGE, BLEU, CodeBERTScore precision, and CodeBERTScore recall to evaluate the correctness of the code.
For ROUGE, a perfect score of one would indicate that all code generated by the model is present in the ground-truth solution. The ROUGE scores measured on the baseline models were at best 0.413. However, the overall scores for the agents did increase slightly compared to the baselines, but the difference was marginal. The only model that was able to increase the score with a notable difference was Qwen 3, with 30 billion parameters, whose baseline model achieved a score of only 0.077, and the agent was able to score 0.343. The score for the baseline means that only a fraction of the generated code matched with the code in the ground-truth solution. But for the agent, 34.3\% of the generated tokens match the ground-truth solutions. 


The BLEU score for the baseline reached 0.229 on Llama 3.3, indicating moderate overlap between the generated output and the ground-truth solutions. However, most baseline models exhibited a weak overlap of less than 0.2. This did not change for the agents, and most models received scores similar to the baselines. The only exception was Qwen 3, with 30 billion parameters, which increased by 0.141, making it the only model that increased by more than 0.1. However, even with the increase, the model can still only be said to have a weak overlap between the generated programs and the ground-truth solutions. 

Both ROUGE and BLEU measured the textual similarity between the generated code and the ground-truth solutions, and access to a compiler did not significantly affect this similarity for most of the models. However, each of the tasks was openly formulated and therefore, multiple solutions can exist for the same task. To ensure that solutions with a low ROUGE and BLEU score does indeed not include the correct solution, the solutions scored lower than 0.05 for both ROUGE and BLEU was manually evaluated on if there was solutions to the given task. However, none of the solutions below this score was functional code for the task. In a majority of the cases, the solutions evaluated solely included textual information without code.

When it comes to semantic similarity, measured with CodeBERTScore precision and CodeBERTScore recall, the agents and baselines score about the same, with only marginal increases or decreases. The measured precision and recall indicated that the generated programs were fairly similar to the ground-truth solutions, but not enough to conclude that the generated code was functionally correct. Therefore, three problems (easy, medium, and hard) were chosen for manual evaluation of functionality between the baselines and agents. 

For the easy problem, Unix-ls, which asks to implement the functionality of the command ls in Unix in C, and present a sorted list of files and directories in a specified path. 100 Prisoners was the medium problem, and Monty Hall was the most challenging problem. The functionality of each task was classified into three categories. Correct: when the solution solves the given task; partly correct: when the generated solution is an apparent attempt to solve the task, but lacks crucial components; incorrect: all generated solutions with compiler error, runtime error, or do not attempt to solve the given problem. 

For all three tasks, the overall incorrect-solution rate is lower for agents than for the baselines, as seen in Table \ref{tab:funcprob}. For the easy problem, the baseline models have seven incorrect solutions, compared to the four for the agents. The same patterns can be found for the medium (ten for baseline compared to five for agents) and hard problem (nine for baseline compared to five for agents). However, the number of correct solutions is not significantly different, indicating that the functionality of the code generated by the agent does not differ considerably from that of the baseline models.

\begin{table}[ht]
\caption{Functionality of generated code for each model for three different tasks. Legend: 
\checkmark=Correct, $\sim$=Partly correct, $\times$=Incorrect. B=Base, A=Agent.}
\label{tab:funcprob}
\centering
\scriptsize
\setlength{\tabcolsep}{9pt}

\begin{tabular}{lcccccc}
\toprule
\textbf{Model} & \multicolumn{2}{c}{\textbf{Easy}} & \multicolumn{2}{c}{\textbf{Medium}} & \multicolumn{2}{c}{\textbf{Hard}} \\
 & \textbf{B} & \textbf{A} & \textbf{B} & \textbf{A} & \textbf{B} & \textbf{A} \\
\midrule
Code Llama 7B & $\sim$ & $\sim$ & $\times$ & $\times$ & $\times$ & $\times$ \\
\hline
Gemma 2 2B & $\sim$ & $\sim$ & $\times$ & $\sim$ & \checkmark & \checkmark \\
\hline
Gemma 3 12B & $\sim$ & $\sim$ & $\sim$ & $\sim$ & $\times$ & \checkmark \\
\hline
Gemma 3 1B & $\times$ & $\times$ & $\times$ & $\times$ & $\times$ & $\sim$ \\
\hline
Gemma 3 27B & $\sim$ & \checkmark & \checkmark & \checkmark & $\times$ & $\sim$ \\
\hline
GPT OSS 20B & \checkmark & \checkmark & \checkmark & \checkmark & \checkmark & $\sim$ \\
\hline
Llama 3.2 1B & $\times$ & $\sim$ & $\times$ & $\times$ & $\times$ & $\times$ \\
\hline
Llama 3.2 3B & $\sim$ & $\times$ & $\sim$ & $\sim$ & $\times$ & $\times$ \\
\hline
Llama 3.3 70B & \checkmark & $\sim$ & \checkmark & \checkmark & $\times$ & $\times$ \\
\hline
Mistral 7B & $\times$ & $\sim$ & $\times$ & $\sim$ & $\times$ & $\sim$ \\
\hline
Phi-4 14B & \checkmark & \checkmark & \checkmark & \checkmark & \checkmark & $\sim$ \\
\hline
Qwen 3 30B & $\sim$ & $\sim$ & $\times$ & \checkmark & $\sim$ & \checkmark \\
\hline
Qwen 3 4B & $\times$ & \checkmark & $\times$ & \checkmark & \checkmark & \checkmark \\
\hline
SmolLM 2 1.7B & $\times$ & $\sim$ & $\times$ & $\times$ & \checkmark & \checkmark \\
\hline
SmolLM 2 135M & $\times$ & $\times$ & $\times$ & $\sim$ & $\sim$ & \checkmark \\
\hline
SmolLM 2 360M & $\times$ & $\times$ & $\times$ & $\times$ & $\times$ & $\times$ \\
\bottomrule
\end{tabular}
\end{table}

The successful compilation rate increased for agents compared to the baselines, without affecting the textual or semantic differences between the generated programs and the ground-truth solutions. This indicates that the agents focus on repairing compiler errors without making changes to the code that are unrelated to these errors.

\begin{table*}[t] 
\noindent
\caption{LMs and Agents. Succ -- Successful Compile Result, R.1 -- ROUGE1 metrics, CB Prec. -- CodeBERT Precision, CB R. -- CodeBERT Recall}
\label{tab:allresults}
\centering
\small
\setlength{\tabcolsep}{5pt}
\begin{tabular}{
  l        
  c        
  S[table-format=3.2] S[table-format=3.2]  
  S[table-format=3.2] S[table-format=3.2]  
  S[table-format=3.2] S[table-format=3.2]  
  S[table-format=3.2] S[table-format=3.2]  
  S[table-format=3.2] S[table-format=3.2]  
}
\toprule
& & \multicolumn{2}{c}{Succ} & \multicolumn{2}{c}{R1} & \multicolumn{2}{c}{BLEU} & \multicolumn{2}{c}{CB Prec} & \multicolumn{2}{c}{CB R.} \\
\cmidrule(lr){3-4}\cmidrule(lr){5-6}\cmidrule(lr){7-8}\cmidrule(lr){9-10}\cmidrule(lr){11-12}
\cmidrule(lr){3-4}\cmidrule(lr){5-6}\cmidrule(lr){7-8}\cmidrule(lr){9-10}\cmidrule(lr){11-12}
Model & Size & {Base} & {Agent} & {Base} & {Agent} & {Base} & {Agent} & {Base} & {Agent} & {Base} & {Agent} \\
\midrule
Code Llama&	7B&	    48,2\%&	    78,8\%&	0,305&	0,303&	0,085&	0,119&	0,731&	0,734&	0,714&	0,712 \\
Gemma 2&	    2B&	    60,4\%&	    72,5\%&	0,319&	0,329&	0,105&	0,094&	0,741&	0,753&	0,716&	0,716 \\
Gemma 3&	    12B&	    84,5\%&	94,7\%&	0,392&	0,398&	0,205&	0,219&	0,772&	0,776&	0,758&	0,76 \\
Gemma 3&	    1B&	32,9\%&	    38,2\%&	0,268&	0,217&	0,083&	0,055&	0,732&	0,702&	0,68&	0,653 \\
Gemma 3&	    27B&	90,3\%&	98,0\%&	0,393&	0,401&	0,228&	0,233&	0,768&	0,775&	0,763&	0,765 \\
GPT OSS&	20B&	84,3\%&	97,7\%&	0,39&	0,396&	0,196&	0,206&	0,752&	0,756&	0,769&	0,773 \\
Llama 3.2&	1B&	38,3\%&	    55,2\%&	0,325&	0,32&	0,146&	0,171&	0,737&	0,724&	0,731&	0,736 \\
Llama 3.2&	3B&	    52,9\%&	    72,1\%&	0,381&	0,384&	0,149&	0,194&	0,768&	0,769&	0,747&	0,751 \\
Llama 3.3&	70B&	    \textbf{92,0}\%&	 \textbf{99,9}\%&	0,413&	0,416&	0,229&	0,233&	0,775&	0,779&	0,769&	0,77 \\
Mistral&	7B&	    48,4\%&	    73,7\%&	0,375&	0,377&	0,184&	0,169&	0,75&	0,758&	0,754&	0,749 \\
Phi-4&	    14B&	    84,7\%&	98,3\%&	0,382&	0,394&	0,224&	0,23&	0,745&	0,755&	0,772&	0,773 \\
Qwen 3&	    30B&	    62,7\%&	95,7\%&	0,077&	0,343&	0,014&	0,155&	0,646&	0,74&	0,78&	0,751 \\
Qwen 3&	    4B&	    \textbf{18,0}\%&	 \textbf{97,4\%}&	0,066&	0,129&	0,023&	0,041&	0,646&	0,648&	0,78&	0,745 \\
SmolLM 2&	1.7B&	    39,2\%&	50,2\%&	0,339&	0,32&	0,171&	0,151&	0,739&	0,729&	0,739&	0,734 \\
SmolLM 2&	135M&	    2,9\%&	12,7\%&	0,148&	0,124&	0,008&	0,007&	0,641&	0,615&	0,644&	0,628 \\
SmolLM 2&	360M&	    7,4\%&	25,5\%&	0,201&	0,188&	0,038&	0,029&	0,677&	0,66&	0,652&	0,652 \\

\bottomrule
\end{tabular}
\label{tab:baseline-agent}
\end{table*}

\begin{figure}[ht]
  \includegraphics[scale=0.22]{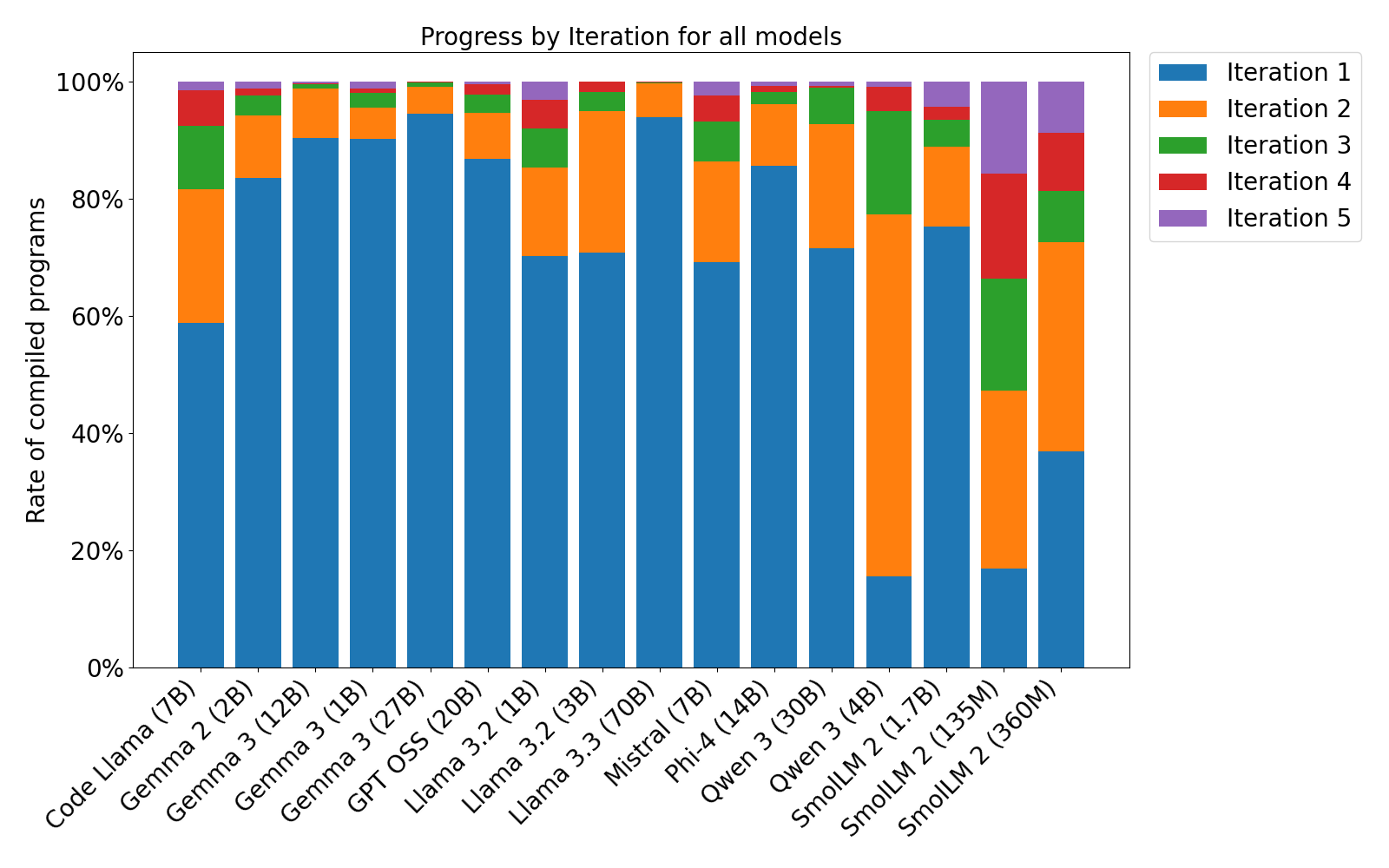}
  \caption{Rate of programs that would successfully compile by all models per iteration.}
  \label{fig:rate by iteration}
\end{figure}

To understand how the iterative process between the LM and the compiler affected the total number of successfully compiled programs, we analyzed the number of programs that compiled successfully for each iteration (see Table \ref{tab:iterations}). A pattern observed in most models was that the positive effect of the increased number of generated programs that compiled decreased with each iteration. By iteration three, all models had generated 92 \% or more of the programs that would successfully compile, except the two smallest SmolLM models, (see Figure \ref{fig:rate by iteration}). Thus, the iterative process has a positive impact, with the effects being most significant for the first few iterations.

\begin{table}[b] 
  \centering
 \caption{}
 \label{tab:iterations}
  
\begin{tabular}{ |p{1.6cm}|p{0.6cm}|p{0.4cm}|p{0.4cm}|p{0.4cm}|p{0.4cm}|p{0.4cm}| }
 \hline
 \multicolumn{7}{|c|}{Successfully compiled programs per iteration} \\
 \hline
 Model&	Size&	1&	2&	3&	4&	5\\
\hline
Code Llama&	7B&	    324&	126&   59&	34&	8 \\
Gemma 2&	    2B&	    425&	55&	   17&	6&	6\\
Gemma 3&	    12B&	598&	56&	   5&	1&	2 \\
Gemma 3&	    1B&	    241&	14&	   7&	2&	3\\
Gemma 3&	    27B&	647&	32&	   5&	1&	0\\
GPT OSS&	20B&	593&	54&	   21&	12&	3\\
Llama 3.2&	1B&	    272&	58&	   26&	19&	12\\
Llama 3.2&	3B&	    357&	122&   16&	9&	0\\
Llama 3.3&	70B&	656&	40&	   1&	1&	0\\
Mistral&	7B&     356&	89&	   35&	23&	12 \\
Phi-4&       14B&	589&	73&	   14&	7&	5\\
Qwen 3&      30B&	\textbf{479}&	\textbf{141}&   42&	2&	5\\
Qwen 3&      4B&	    \textbf{106}&	\textbf{421}&   120&	28&	6\\
SmolLM 2&	1.7B&	264&	48&	   16&	8&	15\\
SmolLM 2&	135M&	15&	    27&	   17&	16&	14\\
SmolLM 2&	360M&	67&     65&	   16&	18&	16\\

 \hline
\end{tabular}
\end{table}

To understand how each model altered the generated output between iterations, we analyzed the ROUGE, BLEU, and the change in number of rows to assess the text similarity between iterations. For all models except SmolLM 2 and Qwen 3, there was high similarity in the generated code across iterations; see Table \ref{tab:model_iter_metrics_singlepage}. This strengthens our previous statement that the models focus on repairing the generated code based on the feedback from the compiler.

In summary, the feedback from the compiler helps the LMs to generate executable programs. However, the effect differs between the models used. The model that was affected the most by the compiler was Qwen 3, with 4 billion parameters. It increased its rate of executable programs by 79.4\% and outperformed all baseline models, ranking among the best-performing agents. This means that this model, with access to a compiler, can be used instead of larger models, such as Llama 3.3, and can run on less resource-heavy equipment like a laptop. 

\begin{tcolorbox}[colback=beige, colframe=beige, boxrule=0pt,
                  left=8pt,right=8pt,top=8pt,bottom=8pt, breakable]
\textbf{Answer to RQ1:} The models increase their success rate by 5.3 to 79.4 percentage points, with a limited effect on the semantic and textual differences in the generated code. Qwen 3 with 4 billion parameters had the largest increase of 79.4 \% success rate and thereby beat all baseline models. 
\end{tcolorbox}

\begin{table*}[!ht]
\centering
\captionsetup{font=small}
\caption{BLEU, R.1 -- ROUGE1 metrics, R.Ch -- Number of rows changed}
\label{tab:model_iter_metrics_singlepage}
\scriptsize
\begin{adjustbox}{max width=\textwidth, max totalheight=\textheight}
\begin{tabular}{rp{1cm}rrrrrrrrrrrrrrr}
\toprule
 &  & \multicolumn{3}{c}{Iteration 1-2} & \multicolumn{3}{c}{Iteration 2-3} & \multicolumn{3}{c}{Iteration 3-4} & \multicolumn{3}{c}{Iteration 4-5} \\
\cmidrule(lr){3-5} \cmidrule(lr){6-8} \cmidrule(lr){9-11} \cmidrule(lr){12-14} 
Model & Size & BLEU & R1 & R.Ch & BLEU & R1 & R.Ch & BLEU & R1 & R.Ch & BLEU & R1 & R.Ch \\
\midrule
Code Llama&	7B& 0.660 & 0.711 & 10.3 & 0.814 & 0.741 & 9.4 & 0.831 & 0.811 & 8.1 & 0.850 & 0.830 & 7.5   \\
Gemma 2&	    2B& 0.877 & 0.915 & 3.1 & 0.859 & 0.926 & 3.7 & 0.938 & 0.961 & 2.2 & 0.935 & 0.964 & 1.9   \\
Gemma 3&	    12B& 0.915 & 0.933 & 5.5 & 0.980 & 0.992 & 0.8 & 0.984 & 0.996 & 0.7 & 0.992 & 0.995 & 0.3   \\
Gemma 3&	    1B& 0.540 & 0.722 & 20.5 & 0.370 & 0.683 & 41.0 & 0.298 & 0.634 & 56.4 & 0.381 & 0.679 & 43.5 \\
Gemma 3&	    27B& 0.908 & 0.936 & 8.9 & 0.913 & 0.952 & 3.1 & 0.997 & 0.997 & 0.1 & 0.998 & 0.999 & 0   \\
GPT OSS&	20B& 0.870 & 0.849 & 13.8 & 0.517 & 0.757 & 34.7 & 0.510 & 0.804 & 42.5 & 0.801 & 0.893 & 6.4   \\
Llama 3.2&	1B& 0.573 & 0.823 & 29.2 & 0.861 & 0.887 & 8.9 & 0.855 & 0.888 & 11.0 & 0.869 & 0.900 & 8.0   \\
Llama 3.2&	3B& 0.782 & 0.914 & 13.9 & 0.887 & 0.948 & 4.3 & 0.950 & 0.968 & 3.7 & 0.961 & 0.972 & 2.3   \\
Llama 3.3&	70B& 0.877 & 0.911 & 7.8 & 0.752 & 0.807 & 18.3 & 0.990 & 0.993 & 0.5 & 0.979 & 0.987 & 1   \\
Mistral&	7B& 0.818 & 0.860 & 9.2 & 0.887 & 0.900 & 5.7 & 0.896 & 0.913 & 5.5 & 0.898 & 0.903 & 5.7  \\
Phi-4&	    14B& 0.925 & 0.946 & 3.9 & 0.952 & 0.969 & 2.9 & 0.941 & 0.959 & 3.0 & 0.958 & 0.965 & 3.1   \\
Qwen 3&	    30B& 0.022 & 0.551 & 112.6 & 0.757 & 0.794 & 9.9 & 0.903 & 0.820 & 3.2 & 0.841 & 0.834 & 4.4   \\
Qwen 3&	    4B& 0 & 0.229 & 338.1 & 0.097 & 0.527 & 97.0 & 0.119 & 0.527 & 116.5 & 0.188 & 0.648 & 107.8   \\
SmolLM 2&	1.7B& 0.530 & 0.845 & 29.7 & 0.780 & 0.887 & 16.3 & 0.744 & 0.834 & 19.5 & 0.702 & 0.800 & 16.5   \\
SmolLM 2&	135M& 0.088 & 0.312 & 92.3 & 0.058 & 0.281 & 104.2 & 0.050 & 0.251 & 121.3 & 0.052 & 0.270 & 103.7  \\
SmolLM 2&	360M& 0.272 & 0.577 & 47.1 & 0.318 & 0.628 & 44.0 & 0.250 & 0.524 & 44.4 & 0.236 & 0.468 & 42.6  \\
\bottomrule
\end{tabular}
\end{adjustbox}
\end{table*}

\subsection{Which patterns can be identified in the tasks that benefit the most when an LM has access to a compiler?}

\begin{figure}[ht]
  \includegraphics[scale=0.35]{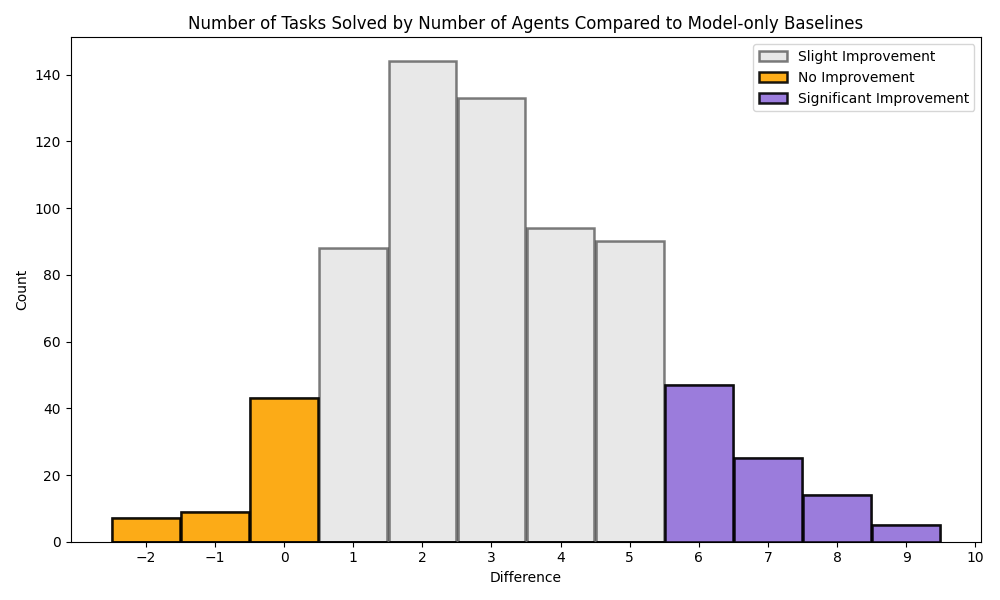}
  \caption{The difference in tasks that the agents generated executable programs for compared to the baselines.}
  \label{fig:DifferenceAgentsVSBaselines}
\end{figure}

Overall, the tasks were solved by the agents with a 5.3 to 79.4 percentage units increase in compiling success, but not all tasks benefited equally. Some tasks were solved by as many agents as baselines. Other tasks saw a moderate increase by the agents, and a few tasks saw a significant rise in the number of models generating solutions that compiled. This suggests that there might be features in the data that the agents can utilize to a higher degree than the baselines. An example of such a task was one called \textit{Prime decomposition} (see Figure \ref{fig:RosettaCodeEx}), for which only 4 of the baselines were able to generate executable programs, but 13 of the agents succeeded. 
We therefore divided the tasks into three categories based on the difference in the number of agents that generated executable solutions to each task compared to the baselines: \textbf{No Improvement}, \textbf{Slight Improvement}, and \textbf{Significant Improvement} (Figure \ref{fig:DifferenceAgentsVSBaselines}).

In tasks that saw a significant improvement in the number of agents solved compared to baselines, we compared the distribution of task categories with the distribution of categories across the entire dataset. However, the distribution remained essentially unchanged, with the broad category "Other Programming Task" followed by "Prime Numbers" as the largest categories. So, the category of the task did not have any significance on the agent's ability to solve the tasks in relation to the baselines.

Even if the categories did not affect the agent's ability to solve the tasks, there could be other patterns in the tasks. We therefore analyzed whether the length of the ground-truth solution could have any relation to how well the agents were able to solve the tasks compared to the baselines. However, the Pearson correlation between the number of tokens in the ground-truth solution and the difference in the number of agents that solved a task compared to baselines was 0.118.  

A similar Pearson correlation (r=0.127) could be found between the number of tokens in the task descriptions and the increased ability of the agents. So, there was a weak correlation, but it is safe to say that neither relation had any significance on the agents' improved ability to solve certain tasks over others.

There might be patterns related to the tasks that explain why the agents improved on specific tasks more than others, but they were not found within the limits of this study. 

\begin{tcolorbox}[colback=beige, colframe=beige, boxrule=0pt,
                  left=8pt,right=8pt,top=8pt,bottom=8pt, breakable]
\textbf{Answer to RQ2:} No pattern could be found in the tasks that would explain why the agents had an improved ability to solve certain tasks over others. The category, length of ground-truth solution, and length of task description were not or very weakly correlated with the agents' increased ability to solve specific tasks.  
\end{tcolorbox}


\subsection{Which patterns can be identified in the errors that benefit the most when an LM has access to a compiler?}

To understand why the tasks in the Significant Improvement category benefited from the agents, we analyzed the types of errors committed compared to the baseline models. The labeling of all programs and error messages was classified, and six categories were identified. 

The first category included all examples where a model generated code in a language other than C, referred to as \textbf{Language Mismatch}. Although the models were all instructed to generate code written in C, there were many instances when this was not followed. 

In some instances, the generated output included no code and consisted either of just a textual description or empty output. These types of errors were categorized into our second category, labeled as \textbf{Missing Code}. However, sometimes the models generated programs in C, but the markdown around the code was missing, and there were textual descriptions (not comments) alongside the code. In these instances, the agents were unable to correctly extract the code because there was no reliable way to distinguish where the code began and ended. The errors caused by this were categorized into our third category \textbf{Markdown Error}.

All errors caused by faulty syntax were categorized into our fourth category \textbf{Syntax Error}. Here, every generated program with a misplaced semicolon, a missing bracket, or other common syntax errors could be found.

Our fifth category was called \textbf{Undefined Reference} and included all cases with calls to undefined functions and missing main functions. For both baselines and agents, about 66\% of these types of errors were because of a missing main function. 

The last category \textbf{Linking Error} included all errors where the primary reason was references to non-existing header files. 

\begin{figure}
  \includegraphics[scale=0.19]{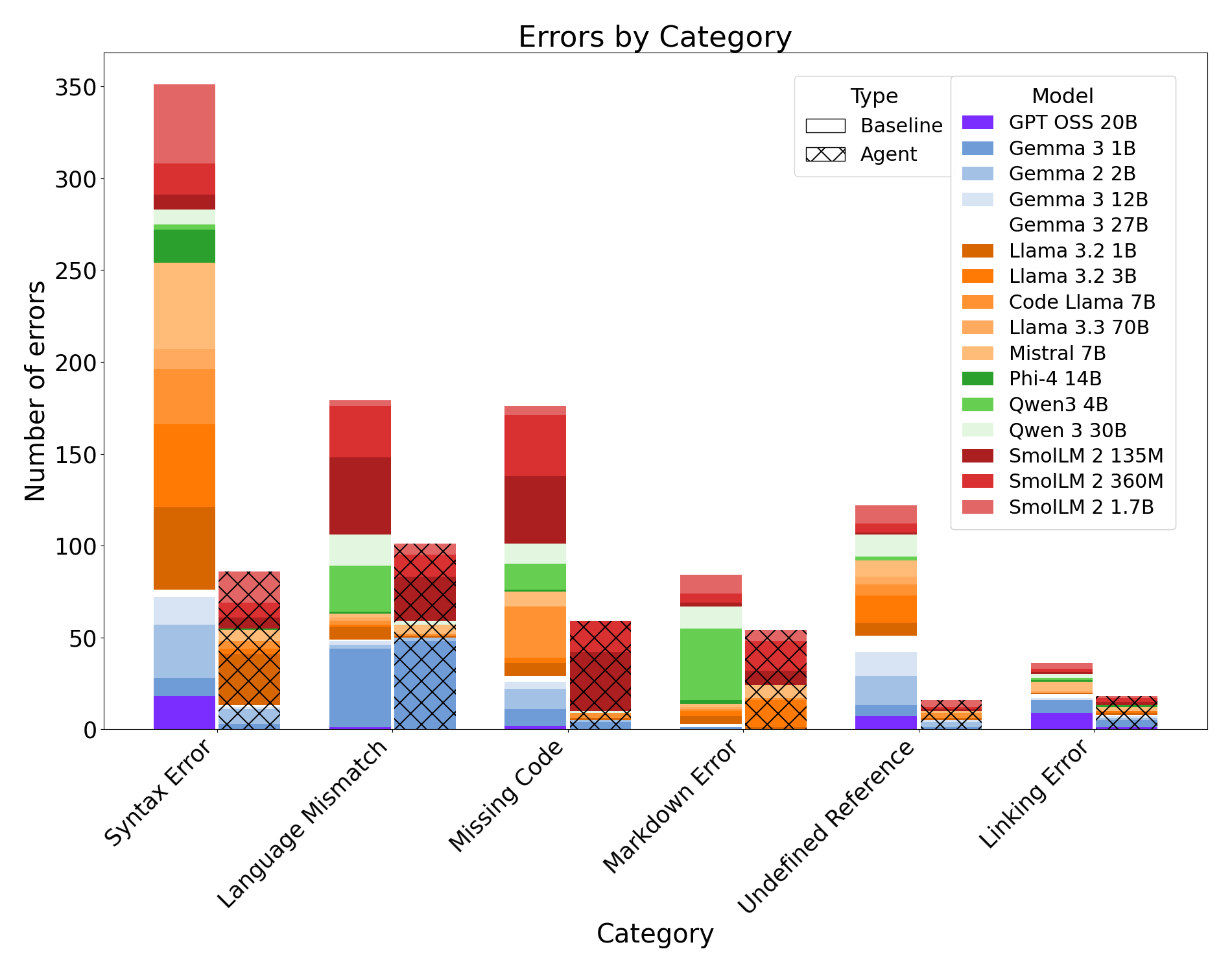}
  \caption{The category of the errors committed by the agents and baselines on the tasks with a significant improvement in executable solutions by the agents.}
  \label{fig:ErrorCatAll}
\end{figure}

We found that the most common error committed by the baselines was Syntax Error, followed by Language Mismatch, Missing Code, Undefined Reference, Markdown Error, and Linking Error, as seen in Figure \ref{fig:ErrorCatAll}. Syntax Errors accounted for around a third of all errors committed by the models. 

The total number of errors that occurred for the agents was lower than that of the baselines, decreasing from 948 erroneous solutions to 334.
Even though all categories of errors decreased, the number of errors related to Undefined References had the most significant decrease, at 86.8\%, followed by Syntax Errors, which saw a reduction of 75.5\%, and Missing Code, with a decline of 66.5\%. The common factor among all three categories was that the error messages could be translated into specific actions needed to repair the code. If a syntax error occurred, the line with the faulty code was clearly stated in the error message. In Undefined References, the message clearly stated which function call was incorrect or references a nonexistent main function. For Missing Code, the error message could be more ambiguous. Still, the agent had access to the conversation history and could deduce that the compiler errors occurred because there was no code.

\begin{figure}
  \includegraphics[scale=0.19]{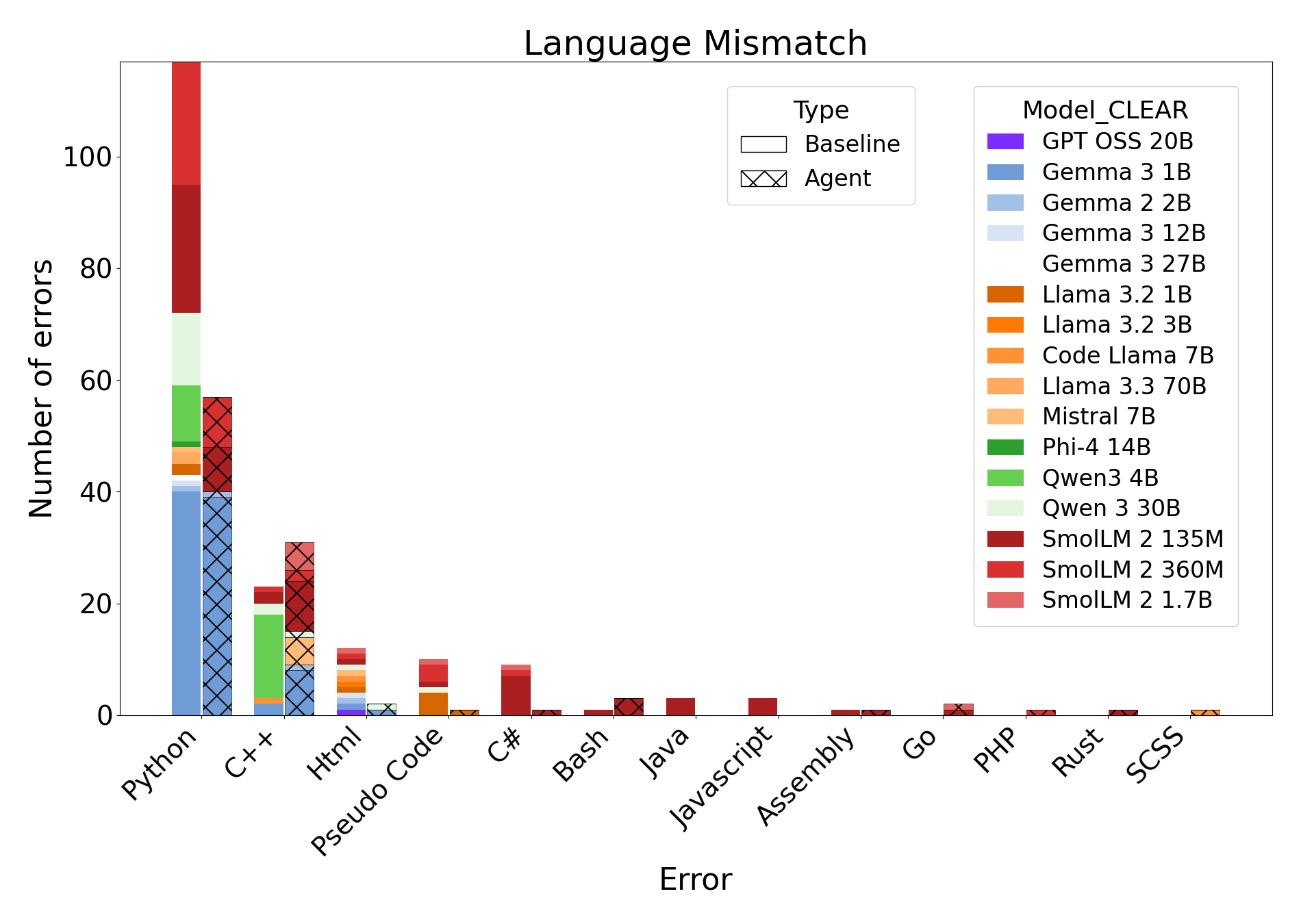}
  \caption{The Programming languages for the generated code with errors related to Language Mismatch for both agents and baselines.}
  \label{fig:langMis}
\end{figure}

The number of errors categorized as Language Mismatch decreased by 43.6 \%.
For both agents and baselines, the most common reason was generated code in Python, followed by C++, as shown in Figure \ref{fig:langMis}.

The category of errors that showed the least significant difference, at 35.7 \%, was the Markdown Error, which accounted for less than 10 \% of the total number of errors for the baselines and 16 \% for the agents. 

One explanation for why both Markdown Error and Language Mismatch were harder for the models could be that the error message was ambiguous, particularly in conjunction with the conversation history.

So, the effect of access to a compiler is also dependent on how well the model can interpret the compiler's error messages. The model will only be able to interpret the error if the error messages are clear and indicate what the issue is in the generated code. Therefore, the effect of a compiler is dependent on the type of compiler and how easily the model can interpret the error messages.

\begin{tcolorbox}[colback=beige, colframe=beige, boxrule=0pt,
                  left=8pt,right=8pt,top=8pt,bottom=8pt, breakable]

\newpage
\textbf{Answer to RQ3:}
 We found six categories of errors in the generated programs: Language Mismatch, Markdown Error, Syntax Error, Undefined Reference, and Linking Error. The agents were able to significantly decrease the errors categorized as Syntax Errors (75\%) and Undefined References (87\%). Both categories had straightforward error messages that clearly indicated the line and error of the faulty code, which the agent could then translate into actions.   
\end{tcolorbox}

\section{Threats to Validity}
\label{sec:threats}

\subsection{External Validity}
Our evaluation relies on RosettaCode as our sole dataset and targets a single programming language, C. This design limits how well we can generalize our findings to other datasets and languages. Although RosettaCode includes a wide range of programming tasks of varying difficulty, we observed no patterns suggesting a relationship between task type and the rate of successful compilations. However, this conclusion is dependent on the predefined categories in RosettaCode. An alternative would be to define the categories manually based on clear definitions, which could yield different patterns. In addition, while C was the target language, the models defaulted to a high degree of Python, indicating a language bias. Therefore, the reported effects should not be assumed to hold over other programming languages. 

RosettaCode consists of single-file programs with a range of functions, but most problems can be solved with only a few (10 or fewer). However, no consideration was given to the number of functions in each program when evaluating the compilation success rate. Therefore, the effect observed in this study is not demonstrated for more complex tasks involving multiple files or a larger number of functions.

\subsection{Construct Validity}
Model behavior is known to be sensitive to the formatting of the prompt, and different models may respond differently to the same prompt. To control this factor, we used the same two prompts for all of our models. This reduces variance in the prompting but introduces a threat to validity: the response from the models may reflect the style of the prompting rather than the models' capabilities.  We leave it to future work to optimize the prompting through model-specific prompt tuning, and more detailed prompts may affect the results.  

The problems on RosettaCode are openly described so that a single problem can be solved in multiple ways. In this study, CodeBERTScore was included alongside ROUGE and BLEU to measure the semantic meaning of the generated code. To ensure that the solutions actually solved each problem, we manually checked the generated solutions with a ROUGE score below 0.05 to determine whether they were solved, even when the similarity between the ground-truth solution and the generated code was low.

\subsection{Internal Validity}
We did not evaluate the functional correctness or runtime behavior of the generated programs. Instead, we compared the semantic similarity of the result to the ground-truth solutions. Semantic similarity is an imperfect proxy for correctness: code can be structurally similar and still incorrect, and correct solutions may diverge semantically from the ground-truth solutions. Therefore, the result may have been different if we had evaluated the code's functionality as well. 

Our agents had five attempts to compile the generated code, while our baseline models had only one attempt. An improved design would be to allow the baselines five attempts to solve the generated code, where they could improve the code from the previous attempt without feedback from the compiler.

\section{Conclusion}
\label{sec:conclusion}

In this paper, we addressed the problem of understanding how access to a compiler affects the performance of models when solving programming tasks. Previously, the improvement of the models was addressed by increasing size, fine-tuning, and providing extra information. However, it can be computationally costly; we need to find more efficient ways that are not as computationally costly to improve performance. 



We addressed this problem by conducting computational experiments with LMs on the community programming dataset RosettaCode, measuring the compilation rate and the textual and semantic similarity to the original solutions. We compared the results from LMs that were allowed to repair the generated programs based on error messages from a compiler, to models without this access to a compiler.


We found that the compilation rate increased by 5.3 to 79.4 percentage points, without affecting the semantic and textual similarity to the original solutions. The model with the most significant increase, Qwen 3, increased the performance from one of the worst (18.0\%) to one of the best (97.4\%). We also found that the models could fix straightforward error messages, such as syntax or undeclared references, but struggled with those containing multiple errors. No clear pattern was found in the task descriptions or the original solutions that would explain the difference in model performance.

\section{Future Work}
Based on our results, we can now establish a model hierarchy and use this hierarchy to further explore systems that begin with a smaller model and transition to a larger one when necessary. This would allow us to use a system that could reduce computational time by using smaller models when eligible. In further work, we plan to include more tools, such as linting tools and software for testing the functionality of the generated programs, which could provide additional information for the model's decision-making. 

Another direction is to examine how well our results generalize to other languages and tasks by including more common languages such as Java, C++, and Python. But also to less common languages that do not have as big a community around them, and therefore less data for the LLMs. As we establish that the effect of the compiler can, to a not-insignificant extent, be attributed to how well the error messages can be interpreted as actions, different compilers for different languages can affect the LLM's ability to solve the error messages generated by those compilers.

\section*{Data Availability}
All data used for evaluation, the results from these evaluations, and the categorization of our data can be found at https://doi.org/10.5281/zenodo.17361190. The repository with the code used can be accessed at https://anonymous.4open.science/r/ChatAI-845D/.

\section*{Acknowledgment}
The Swedish Research Council partially funded this study under grant number 2024-04687. 

\printbibliography 

\end{document}